\documentclass[prb,reprint,aps,amsmath,amssymb,twocolumn,superscriptaddress,longbibliography]{revtex4-2}

\usepackage[utf8]{inputenc}
\usepackage{amsmath}
\usepackage{graphicx}
\usepackage{bm}
\usepackage{mathbbol}
\usepackage{natbib}
\usepackage{hyperref}
\usepackage{comment}
\usepackage[normalem]{ulem}

\usepackage{xcolor}
\definecolor{darkblue}{rgb}{0.,0.,0.4}
\definecolor{darkred}{rgb}{0.5,0.,0.}
\definecolor{BlueViolet}{RGB}{138,43,226}
\definecolor{SkyBlue}{RGB}{30,144,255}
\definecolor{DarkGreen}{RGB}{0,100,0}

\begin{document}
	
	\title{ Conformal Nature of Quantum Phase Transitions via Fuzzy Three-Sphere Regularization }

	\author{Xue Meng}
	\affiliation{Department of Physics, Fudan University, China}
	\affiliation{Department of Physics, School of Science, Westlake University, Hangzhou 310030, China }	

    \author{Liangdong Hu} \email{huliangdong@hnu.edu.cn}
    \affiliation{School of Physics and Electronics, Hunan University, Changsha 410082, China}
    
	\author{W. Zhu} \email{zhuwei@westlake.edu.cn}
	\affiliation{Department of Physics, School of Science, Westlake University, Hangzhou 310030, China }

	\begin{abstract}
		Conformal field theory (CFT) offers a modern viewpoint for understanding phase transitions. However, directly accessing the conformal algebra and microscopically uncovering the emergent conformal symmetry, especially in higher dimensions, remains a significant challenge.  
        Motivated by recent advances in revealing CFT features via the fuzzy two-sphere, here we  generalize this approach to higher dimensions and aim to expose the conformality at the $(3+1)$-D quantum critical point. 
        We demonstrate this framework by investigating quantum phase transitions belonging to the Ising and Yang-Lee universality classes in a $(3+1)$-D quantum model (equivalent to a classical four-dimensional system), realized via Landau level projection on the fuzzy three-sphere. By computing the energy spectra at criticality, we explicitly verify the state-operator correspondence, a hallmark of conformal invariance. Together with prior advances, this work establishes a new pathway for the microscopic study of emergent conformality in higher-dimensional phase transitions.
	\end{abstract}
	
\maketitle

\section{Introduction}

The transitions among distinct phases of matter are ubiquitous\cite{Cardy:1996xt,Sachdev_book,Domb_book}. 
Many ingredients influence the nature of phase transitions, e.g., one of the most widely studied is the underlying symmetry. 
Particularly, a more profound, but often experimentally constrained, factor is the space-time dimension $d$, which directly controls the effective degrees of freedom in the physical system.
Variations in $d$ can lead to qualitatively distinct critical behaviors. 
For instance, the Mermin-Wagner theorem dictates that magnetic systems with continuous symmetry can only exhibit a finite-temperature phase transition in dimensions $d \ge 2$. Another example is 
the extended-to-localization transition, which does not occur in $d=1$ but manifests itself in $d=3$. 

Regarding the theory of phase transitions, conformal field theory (CFT) is a modern framework\cite{DiFrancesco:1997nk,Christe_1994}. CFT allows us to access universal critical exponents and correlation functions in a non-perturbative way. The application of CFT is particularly potent in $d=2$ statistical models\cite{BELAVIN1984333,Polyakov:1970xd}, where the conformal algebra imposes an extraordinarily restrictive set of constraints and enables a wealth of exact results. Nevertheless, the efficacy of CFT diminishes significantly in higher dimensions ($d > 2$), i.e., the conformal algebra produces far fewer analytical constraints, making analytic solutions less tractable. Thus, performing an empirical study of higher-dimensional phase transitions is demanding, and the formidable challenge lies in the heavy computational cost. Very recently, a novel approach has been proposed to revitalize the study of $d=3$ CFTs \cite{PhysRevX.13.021009, 10.21468/SciPostPhys.19.3.076,PhysRevB.110.115113, 
Hu2024, 10.21468/SciPostPhys.17.1.021, 10.21468/SciPostPhys.18.1.031, dedushenko2024isingbcftfuzzyhemisphere,
dey2025conformaldatao3wilsonfisher,x1qn-x6xb,PhysRevX.14.021044, PhysRevLett.132.246503, PhysRevB.110.125153,he2025freerealscalarcft, taylor2025conformalscalarfieldtheory,Fan:2025bhc,EliasMiro:2025msj,ArguelloCruz:2025zuq,10.21468/SciPostPhys.15.6.243,Voinea:2024ryq,Zhou:2024zud,Zhou:2025rmv,Fardelli:2024qla,Fan:2024vcz}. This strategy involves placing the CFT on the geometry of a two-sphere, a configuration that enjoys the state-operator correspondence \cite{JLCardy_1984,JLCardy_1985} and allows the extraction of conformal  data with quite low computational cost. In this context, a natural and compelling intuition arising from this development is to inquire whether a similar strategy of employing a spherical background can be successfully generalized to study CFTs in higher-dimensions $d>3$, potentially opening a new avenue for taming their intricacies.

\begin{figure}[t]
\includegraphics[width=0.45\textwidth]{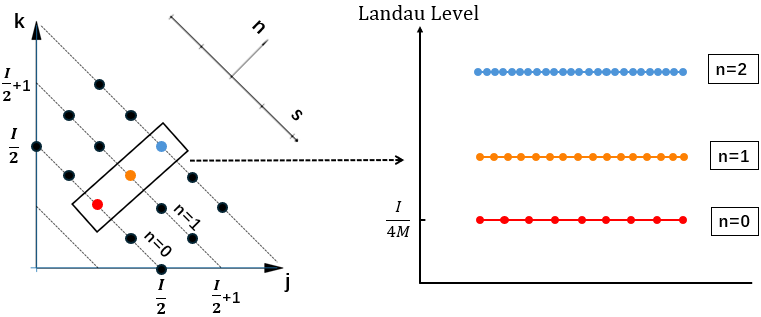}   
\includegraphics[width=0.45\textwidth]{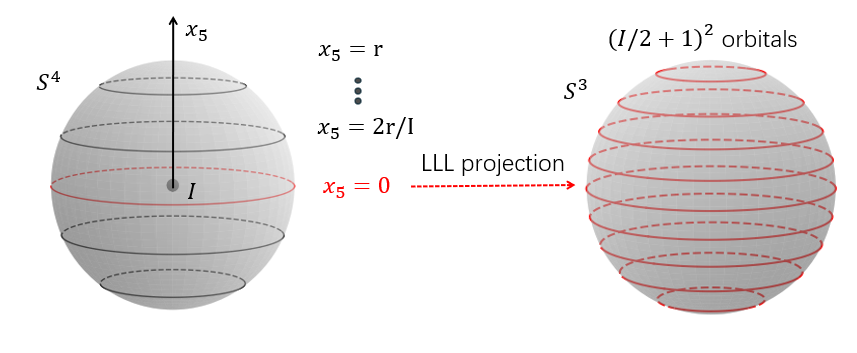} 
\caption{ \textbf{Landau levels of fuzzy three-sphere $\mathbb{S}^3$.}
(Top) Each Landau level on fuzzy $\mathbb{S}^4$ geometry corresponding to a $SO(5)$ irreducible representation \cite{10.1063/1.523618,doi:10.1126/science.294.5543.823}, which can be decomposed into blocks of $SO(4)$ representations labeled by two integer numbers $(n,s)$ \cite{HASEBE2018149}. Taking $s=0$ gives the $SO(4)$ Landau levels on fuzzy $\mathbb{S}^3$.
(Bottom) For $n=0$, the fuzzy $S^3$ locate along $x_5$-axis on $\mathbb{S}^4$  with equal spacing. In particular, $x_5=0$ latitude (red dashed line) on $\mathbb{S}^4$ corresponds to the $n=0,s=0$ $SO(4)$ representation, i.e., the lowest $SO(4)$ Landau level of $\mathbb{S}^3$.}
\label{fig:rep&fuzzy}
\end{figure}

This paper extends the fuzzy sphere approach \cite{PhysRevX.13.021009} and explores a practical path to investigate four-dimensional ($d=4$) phase transitions. The key is to utilize the Landau level projection on fuzzy three-sphere $\mathbb{S}^3$ to regularize the theory and promote the conformal state-operator correspondence. In the examples of the Ising and Yang-Lee transitions, we demonstrate the approximate conformal tower structure of low-lying operators, which reflects the emergent conformal symmetry. A similar strategy can be generalized to fuzzy $\mathbb{S}^{d-1}$-sphere to investigate $d$-dimensional phase transitions. 

This paper is organized as follows. In Sec.\ref{sec:2}, we review the construction of the $SO(4)$ Landau model on the three-sphere. In Sec.\ref{sec:3}, we formulate these Landau levels as a regularization scheme for the 4D Ising and Yang-Lee transitions on a fuzzy three-sphere. In Sec.\ref{sec:4.1}, we employ conformal perturbation theory to determine the Ising phase transition point numerically. The low-lying energy spectra and the associated state-operator correspondence analysis are presented in Sec.\ref{sec:4.2}. In Sec.\ref{sec:4.3} we apply the same methods to study the Yang-Lee transition. Summary and discussion are provided in Sec.\ref{sec:4}.

    \section{Fuzzy 3-sphere and $SO(4)$ Landau levels}
    \label{sec:2}
    To access the conformal data of the theory, we employ radial quantization, in which the quantum Hamiltonian is defined on the $S^{d-1} \times \mathbb{R}$ geometry (i.e., the spatial part of the Hamiltonian lives on a $(d-1)$-dimensional sphere).
    In particular, the eigenstates of the CFT Hamiltonian on $\mathbb{S}^{d-1}\times R$ are in one-to-one correspondence with CFT operators, known as the state-operator correspondence\cite{JLCardy_1984,JLCardy_1985}, establishing a direct link between the energy spectrum $\{E_n\}$ and the scaling dimensions of operators $\Delta_n$: $\delta E_n = E_n - E_0 \propto \Delta_n$. For $d=3$ case, $\mathbb{S}^{2}\times R$ has been successfully realized via the fuzzy two-sphere \cite{PhysRevX.13.021009}. Here we generalize this idea to the fuzzy three-sphere.

    \subsection{The $SO(5)$ Landau level}
    \label{sec:2.1}

    In this section, we first review the $SO(5)$ Landau problem defined on $\mathbb{S}^4$ with Yang's $SU(2)$ monopole gauge field\cite{doi:10.1126/science.294.5543.823,10.1063/1.523506,10.1063/1.523618,Hasebe:2010vp,HASEBE2014952,HASEBE2020115012,PhysRevD.105.065010,PhysRevD.108.126023}, and then adopt the dimensional reduction approach to the $SO(4)$ Landau model on $\mathbb{S}^3$ in Sec.\ref{sec:2.2}.
    Let us consider the real space four-sphere $\mathbb{S}^4$ with coordinates $x_{a=1,2,3,4,5}$ satisfying $x_a x_a = r^2$. Inserting an $SU(2)$ monopole with charge $I$ at the center of $\mathbb{S}^4$ leads to the quantized Landau levels. 
     Since $S^4 \simeq SO(5)/SO(4)$, the 2nd Hopf map $S^7 \overset{S^3}{\to} S^4$ provides a natural description between the $SU(2)$ monopole and the fuzzy four-sphere by introducing a four-component complex spinor $\psi$ living on $S^7$ with the constraint $\psi^{\dagger} \psi = 1$: 
	\begin{equation}
		\psi \to x_a = \psi^{\dagger}\gamma_a \psi,
		\label{eq:2ndHopf}
	\end{equation}
	where $\gamma$ denote the $4 \times 4$ $SO(5)$ gamma matrices satisfying the Clifford algebra $ \left \{ \gamma_a,\gamma_b \right \} = 2 \delta_{ab}$.
	For $r=1$, one can choose an explicit solution of Eq.(\ref{eq:2ndHopf}), except for the south pole, expressed as 
	\begin{equation}
		\psi = \frac{1}{\sqrt{2(1+x_5)}} \binom{(1+x_5) \phi}{(x_4 - i \sigma_i x_i) \phi}, 
		\label{eq:SO5spinor}
	\end{equation}	
	with the two-component complex spinor satisfying $\phi^{\dagger} \phi=1$ and $\sigma_{i=1,2,3}$ the Pauli matrices. The corresponding fibre-connection $A=-i \psi^{\dagger} d\psi$ is then derived as
	\begin{equation}
		A = \phi^{\dagger}dx_a A_a \phi, 
	\end{equation}	
	with the $SU(2)$ gauge field 
	\begin{align}
		A_{\mu=1,2,3,4} &= -\frac{1}{1+x_5}\eta_{\mu \nu i}x_{\nu}S_i, \quad A_5 = 0, \nonumber \\
		\eta_{\mu \nu i} &= \epsilon_{\mu \nu i4} + \delta_{\mu i} \delta_{\nu 4} - \delta_{\mu 4} \delta_{\nu i},
		\label{eq:gaugeA}
	\end{align}	
	where $S_i = \sigma_i / 2$ are the $SU(2)$ isospin matrices of the spin-$I/2$ representation, with $S_i^2 = \frac{I}{2}(\frac{I}{2}+1)$. Here $I$ is an integer denoting the $SU(2)$ monopole charge. Under an $SU(2)$ gauge transformation $ \psi \to \psi' = \psi g, g \in SU(2)$, we have $A \to A'=g^{\dagger}Ag - ig^{\dagger}dg$, thereby defining a nontrivial $SU(2)$ bundle over $S^4$. The choice $g = \frac{1}{\sqrt{1-x_5^2}}(x_4 + ix_i \sigma_i)$ serves as the transition function connecting the northern and southern patches, in analogy with the $U(1)$ bundle over $S^2$ case.
    
	The single-particle Hamiltonian for the QHE problem on $S^4$ with radius $r$ (i.e.,$A \to r^{-1}A(x/r)$) under this $SU(2)$ gauge field is given by
	\begin{equation}
		H = \frac{1}{2Mr^2}\sum_{a<b}{\Lambda_{ab}^2}, 
	\end{equation}		
	where $\Lambda_{ab} = -i (x_aD_b - x_bD_a)$ is the covariant angular momentum and $D_a = \partial_a + iA_a$ is the covariant derivative. The field strength $F_{ab}=-i[D_a,D_b] = \partial_aA_b - \partial_bA_a +i[A_a,A_b]$ can be evaluated as 
	\begin{align}
		F_{\mu \nu} &= \frac{1}{r^2}( -x_{\mu}A_{\nu} + x_{\nu}A_{\mu} + \eta_{\mu \nu i}S_i), \nonumber \\
		F_{\mu 5} &= \frac{1}{r^2}(r+x_5)A_{\mu}.
		\label{eq:fieldF}
	\end{align}	
	The covariant angular momentum $\Lambda_{ab}$ does not form a closed algebra, however, one can construct the $SO(5)$ conserved angular momentum by
	\begin{equation}
		L_{ab} = \Lambda_{ab} + r^2F_{ab}, 
	\end{equation}		
	which generates the $SO(5)$ irreducible representations labeled by two integers $(p,q)_5$ and satisfies the $SO(5)$ commutation relation $[L_{ab},L_{cd}] = i(\delta_{ac}L_{bd} - \delta_{bc}L_{ad}-\delta_{ad}L_{bc}+\delta_{bd}L_{ac})$.
	Since the field strength is radially distributed and the system respects the $SO(5)$ rotational symmetry, one can show the orthogonality condition $\Lambda_{ab}F_{ab} = F_{ab}\Lambda_{ab}=0$ and rewrite the Hamiltonian as
	\begin{equation}
		H = \frac{1}{2Mr^2}\sum_{a<b}{(L_{ab}^2-r^4F_{ab}^2)}.
		\label{eq:Ham5}
	\end{equation}	
	For an $SO(5)$ irreducible representation $(p,q)_5$, the first term in Eq.(\ref{eq:Ham5}) corresponds to the Casimir operator $C_2(p,q)$, which reads as $ \sum_{a<b}{L_{ab}^2} = \frac12p^2 + \frac12q^2 +2p +q$. The corresponding dimension is $d(p,q) = \frac16(q+1)(p+2)(p+q+3)(p-q+1)$.  The second term is $\sum_{a<b}{F_{ab}^2} = \sum_{\mu<\nu}{(\eta_{\mu \nu i}S_i)^2} = 2S_i^2 = \frac12I(I+2)$. So the energy of this QHE model is nothing but the eigenvalue of the $SO(5)$ Casimir operator up to an additive constant.
    
	Since $SO(4)$ is a subgroup of $SO(5)$ generated by six of the ten $SO(5)$ generators, two commuting $SU(2)$ algebras generated by $\left \{  J^2,J_3\right \}$ and $\left \{K^2,K_3 \right \}$ can be constructed. These $SO(4) \simeq SU(2)_L \otimes SU(2)_R$ degrees of freedom are labeled by the corresponding $SU(2)$ quantum numbers $\left \{  j,m_j,k,m_k \right \}$. From the viewpoint of group representation theory, one may expect that the eigenstates which simultaneously diagonalize $\left \{ C_2(p,q), J^2,J_3,K^2,K_3 \right \}$ form a basis for an irreducible $SO(5)$ representation. In the presence of a given $SU(2)$ monopole background $I$, Yang shows those representations $p$ and $q$ are related by $p-q=I$.
    
	Thus, the eigenvalues of each quantized Landau level correspond to an $SO(5)$ irreducible representation $(p=q+I,q)_5$ and can be written as
	\begin{equation}
		E(p=q+I,q) = \frac{1}{2Mr^2}(q^2 + q(I+3) + I),
	\end{equation}
	where $q=0,1,2,\dots$ denotes the $SO(5)$ Landau level index. The irreducible representation $(p,q)_5$ of the $N^{th}$ Landau level can be decomposed into blocks of $SO(4)$ irreducible representations $(j,k)_4$ as
	\begin{align}
		(p=N+I,q=N)_5 = \sum (j,k)_4  \nonumber \\
		\equiv \sum^N_{n=0} \sum^{\frac I2}_{s=-\frac I2} (\frac{n+s}2+\frac I4, \frac{n-s}2+\frac I4)_4,
		\label{eq:decomposition}
	\end{align}
	where $n=j+k-\frac I2=0,1,\dots,N$ labels the internal $SO(4)$ Landau level index, which we will explain later. The quantity $s=j-k=\frac I2,\frac I2-1,\dots,-\frac I2$ characterizes the difference between $j$ and $k$, and is referred to as the chirality parameter. 
    The corresponding eigenstates within each Landau level, $\Psi_{N;j,m_j,k,m_k}$, can also be obtained by solving the differential equation using the method of separation of variables (see ref\cite{10.1063/1.523618,HASEBE2020115012}), with polar coordinates on the four-sphere corresponding to the $SO(5)$ monopole harmonics
	\begin{equation}
		\Psi_{N;j,m_j,k,m_k}(\Omega_4) = G_{N,j,k}(\xi)Y_{j,m_j,k,m_k}(\Omega_3),
		\label{eq:wavefun1}
	\end{equation}
	where $Y_{j,m_j,k,m_k}$ is the $SO(4)$ part describing the $S^3$-latitude, which is related to the $SO(4)$ spherical harmonics, and $G_{N,j,k}(\xi)$ is the remaining azimuthal part(i.e.,$x_5 = rcos\xi$). In particular, from a group theory point of view and by using Eq.(\ref{eq:SO5spinor}), one can express the eigenfunction of the LLL in a simple but equivalent form as
	\begin{equation}
		\Psi_{LLL} = \sqrt{\frac{I!}{m_1!m_2!m_3!m_4!}}\psi_1^{m_1}\psi_2^{m_2}\psi_3^{m_3}\psi_4^{m_4},
		\label{eq:wavefun2}
	\end{equation}
	where $m_{i=1,2,3,4}$ are non-negative integers satisfying $m_1+m_2+m_3+m_4 = I$. The relations between the quantum numbers in Eq.(\ref{eq:wavefun2}) and Eq.(\ref{eq:wavefun1}) (with $N=0$) are given by $j=\frac12(m_1+m_2), m_j = \frac12(m_1-m_2), k=\frac12(m_3+m_4), m_k=\frac12(m_3-m_4)$. The degeneracy of the LLL is given by $d_{LLL} = \frac16(I+1)(I+2)(I+3)$.

    \subsection{The $SO(4)$ Landau level}
    \label{sec:2.2}
    The decomposition of those irreducible $SO(5)$ representations from Eq.(\ref{eq:decomposition}) implies a nested structure and provides a natural framework for constructing the lowest Landau level problem on $S^3$ \cite{Hasebe2022QuantumMG,HASEBE2014681,HASEBE2017475,HASEBE2018149,HASEBE2020115012,DEALFARO1976163,deAlfaro:1978dz,Nair:2003st}. To see this, we introduce the fuzzy geometry and rewrite the coordinates $x_a$. In the lowest Landau level (LLL) $q=0$, the kinetic energy is quenched and the Lagrangian is reduced to $L_{LLL} = i \psi^{\dagger} \frac d{dt} \psi$. Thus, by canonical quantization, the orbital coordinates $x_a$ are represented by operators
	\begin{equation}
		X_a = \frac rI \psi^{T}\gamma_a \frac{\partial}{\partial \psi}.
	\end{equation}
    This operator representation of the orbital coordinates naturally leads to noncommutative geometry, which defines a fuzzy manifold emerging from the lowest Landau level physics. This representation also implies that the LLL wavefunction is an eigenstate of $X_5$, leading to quantized eigenvalues of $x_5$.
    Thus, for the lowest Landau level, the degenerate Landau orbits are distributed along the fifth direction by taking the mean values of the coordinate $\langle x_5\rangle = \frac {2r}I s$ with $s=-\frac I2,\dots,\frac I2$ on $\mathbb{S}^4$. 
    Therefore, the fuzzy $\mathbb{S}^3$ is well-defined at a fixed latitude along the $x_5$-axis with position $x_5 = -r, -r+\frac {2r}I, \dots, r$. 
    
    One way to construct the $SO(4)$ Landau model is to perform a dimensional reduction from the higher-dimensional formulation by restricting to the $S^3$ coordinates $x_{a=1,2,3,4}$ only(i.e.,$x_5 \to 0$). In this sense, the so-called meron gauge field is obtained from Eq.(\ref{eq:gaugeA}) as \cite{DEALFARO1976163,deAlfaro:1978dz}
	\begin{equation}
		A_{\mu=1,2,3,4}^{meron} = -\frac{1}{r^2} \eta_{\mu \nu i}x_{\nu}S_i,
		\label{eq:merongauge}
	\end{equation}
	and the corresponding field strength becomes $F_{\mu \nu} = \frac{1}{r^2}( -x_{\mu}A_{\nu}^{meron} + x_{\nu}A_{\mu}^{meron} + \eta_{\mu \nu i}S_i)$.
	Following the same step as in Sec.\ref{sec:2.1}, the covariant and total $SO(4)$ conserved angular momentum $\Lambda_{ab}$ and $L_{ab} =\Lambda_{ab} + r^2F_{ab}$ can be defined respectively. By introducing the $SU(2)_L$ and $SU(2)_R$ operators
	$J_i = -i \frac12 \eta_{\mu \nu i}x_{\mu} \partial_{\nu}$ and $K_i = -i \frac12 \bar{\eta}_{\mu \nu i}x_{\mu} \partial_{\nu} + S_i$,	
	one can rewrite the $SO(4)$ Casimir as $\sum_{a<b} {L^2_{ab}} = 2(J^2+K^2)$. Therefore, the $SO(4)$ Landau Hamiltonian can be expressed as
	\begin{align}
		H &= \frac1{2Mr^2} \sum_{a < b}\Lambda_{ab}^2 = \frac1{2Mr^2} \sum_{a < b}L_{ab}^2 - r^4F_{ab}^2 \nonumber \\ 
		& =\frac1{2Mr^2}(2J^2+2K^2-S^2).
	\end{align}	
    As the $SO(4)$ irreducible representation $(j,k)_4$ can be equivalently labeled by $(n,s)$ defined in Eq.(\ref{eq:decomposition}), in general, for a given $SO(5)$ Landau level, there are $(I+1)$ $\mathbb{S}^3$-latitudes along the $x_5$-axis, 
    each of which has different $SO(4)$ Landau levels labeled by index $n$ with eigenvalues derived as 
    \begin{equation}
	E_n(s) = \frac{1}{2Mr^2}[n(n+2)+\frac I2(2n+1)+s^2].
	\label{eq:so4Energy}
    \end{equation}
    where $n=0,1,\dots,q$.
    Fig. \ref{fig:rep&fuzzy}(bottom) shows the Landau orbits on $\mathbb{S}^4$ and the corresponding fuzzy $\mathbb{S}^3$-latitude for $n=0$.      

    In particular, we focus on the LLL of the fuzzy $\mathbb{S}^3$,
    i.e., $s=0$ and $n=0$, which is located at latitude $\langle x_5\rangle =0$ on $\mathbb{S}^4$
    (marked in red in Fig. \ref{fig:rep&fuzzy}(bottom)). Its energy is $E_{0} = \frac{I}{4 M r^2}$ and has a finite degeneracy as $\mathcal{D} =(I/2+1)^2$. 
    These degenerate Landau orbits form the $SO(4)$ representation $(j,k)_4 = (\frac I4, \frac I4 )_4$, whose eigenfunctions can be explicitly written as tensor products of two spinor components. This structure can be alternatively realized by introducing an $SU(2)$ monopole gauge on $S^3$, given by
    \begin{equation}
	A_{i=1,2,3} = -\frac{1}{r(r+x_4)} \epsilon_{ijk}x_jS_k, \quad A_4 = 0,
	\label{eq:NDgauge}
	\end{equation}
	which was first proposed in ref\cite{Nair:2003st} and later discussed in ref\cite{HASEBE2014681,HASEBE2017475,HASEBE2018149}, and the corresponding field strength is
    \begin{align}
	F_{ij} &= \frac{1}{r^2}( -x_iA_j + x_jA_i + \epsilon_{ijk}S_k), \nonumber \\
	F_{i4} &= \frac{1}{r^2}(r+x_4)A_i.
	\label{eq:NDfield}
	\end{align}	
	In this framework, the gauge field exhibits a string-like singularity analogous to that in the $SO(5)$ Landau problem, similar to Yang's monopole or the Dirac monopole\cite{dirac1931quantised}. While the meron gauge field in Eq.(\ref{eq:merongauge}) has a point-like singularity similar to that of the 3d Wu-Yang $SU(2)$ monopole\cite{wu1976dirac}, the construction in Eq.(\ref{eq:merongauge}) and Eq.(\ref{eq:NDgauge}) provides a four-dimensional generalization of the Wu-Yang and Dirac monopoles, respectively. 
    Following the same steps, the Hamiltonian can be written in terms of the $SO(4)$ Casimir with $SU(2)_{L,R}$ quantum numbers, yielding the eigenvalues in Eq.(\ref{eq:so4Energy}). The equivalence implies a singular $SU(2)$ transformation relating the gauge fields. Further details on the dimensional-hierarchy construction are given in ref\cite{HASEBE2020115012}.

	Further, the similarity between Eq.(\ref{eq:NDgauge},\ref{eq:NDfield}) and Eq.(\ref{eq:gaugeA},\ref{eq:fieldF}) suggests considering a reduced Hopf map instead of the $2^{nd}$ Hopf map to describe the LLL on odd-dimensional spheres, namely the chiral Hopf map
	\begin{equation}
		S^3_L \otimes S^3_R \overset{S^3_D}{\to} S^3.
	\end{equation}
    In this construction, we introduce two chiral Hopf spinors $\psi_L$ and $\psi_R$, which are two-component complex spinors parametrizing $S^3_L \otimes S^3_R$. The normalization condition gives $\psi^{\dagger}_L \psi_L = \psi^{\dagger}_R \psi_R = \frac12$. The chiral Hopf map can be written as \cite{HASEBE2014681,HASEBE2017475}
	\begin{equation}
		\psi_L, \psi_R \to x_{\mu} = \psi^{\dagger} \gamma_{\mu} \psi = 
		\psi^{\dagger}_R q_{\mu} \psi_L + \psi^{\dagger}_L \bar{q}_{\mu} \psi_R,
		\label{eq:chiralHopf}
	\end{equation}
	where $x_{\mu=1,2,3,4}$ satisfies $x_{\mu}x_{\mu}=1$ on $S^3$ and $\gamma_{\mu}$ are now the $SO(4)$ gamma matrices. The second equation we use the quaternions instead of gamma matrices with definition $q_{\mu} = (q_i,q_4) = (-i\sigma_i,1)$, $\bar{q}_{\mu} = (i\sigma,1)$. In this form, one can choose the explicit spinor representation as
	\begin{equation}
		\psi = \binom{\psi_L}{\psi_R} = 
		\frac{1}{\sqrt{2(1+x_4)}} \binom{(1+x_{\mu}q_{\mu}) \phi}{(1+x_{\mu}\bar{q}_{\mu}) \phi}, 
		\label{eq:SO4spinor}
	\end{equation}		
	where $\phi$ is a two-component complex spinor representing the $S^3_D$-fibre. The corresponding $SU(2)$ connection is obtained as $A=-i\frac12(\psi^{\dagger}_Ld\psi_L + \psi^{\dagger}_Rd\psi_R)$, which reproduces Eq.(\ref{eq:NDgauge}). The eigenstates of the LLL can thus be identified with the chiral Hopf spinors, forming the $(j,k)_4 = (\frac I4, \frac I4)_4$ representation:
	\begin{equation}
		\Psi_{n=0,s=0} = \Psi_{j,m_j}^L \otimes \Psi_{k,m_k}^R,
		\label{eq:wavefun3}
	\end{equation}	
	where
	\begin{align}
		\Psi_{j,m_j}^L &= \frac1{\sqrt{(j+m_j)!(j-m_j)!}} \psi_{L_1}^{j+m_j} \psi_{L_2}^{j-m_j}, \nonumber \\
		\Psi_{k,m_k}^R &= \frac1{\sqrt{(k+m_k)!(k-m_k)!}} \psi_{R_1}^{k+m_k} \psi_{R_2}^{k-m_k},
	\end{align}	
	with $m_j = -j, -j+1, \dots,j$ and $m_k = -k, -k+1, \dots, k$. Here, $\psi^{L} = (\psi_{L_1} \psi_{L_2})$ and $\psi^{R} = (\psi_{R_1} \psi_{R_2})$ are two-component chiral Hopf spinors.
    These degenerate Landau orbitals form a finite Hilbert subspace, and they are separated from the higher Landau levels by a finite energy gap (Fig. \ref{fig:rep&fuzzy}(top)). Thus, one can project into the LLL subspace to construct an effective theory and safely project out the higher Landau levels, referred to as the LLL projection\cite{PhysRevLett.51.605,Girvin1990}. Importantly, this projection process truncates the Hilbert space to a finite size but preserves the spatial rotational symmetry of the spherical geometry.

\begin{figure}[b]
\includegraphics[width=0.45\textwidth]{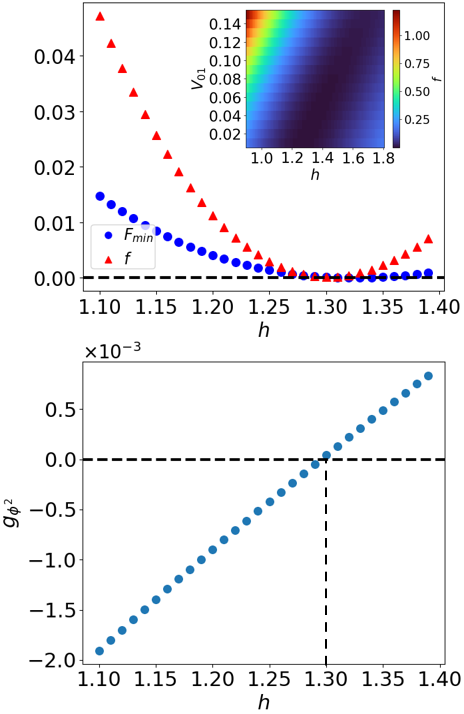} 

\caption{ \textbf{Locating the critical point}.
(Top) Cost functions $f$ (Eq. \ref{eq:costfun}) and $F_{\min}$ (Eq. \ref{eq:costfun2}) as a function of transverse field $h$. 
(Bottom) In the perturbative theory, the coupling $g_{\phi^2}$ which minimizing the cost function $F_{\min}$ at each $h$. The calculations are performed for the system size $N=9$. For the potential $V_{0,0}=1.2, V_{0,1}=V_{1,0}=0.02$, the critical point is determined to be $h_c\approx 1.3$. 
(Inset) Heatmap of $f(h,V_{0,1})$ by setting $V_{0,0}=1.2$.
}
\label{fig:CPT}
\end{figure}

\section{MODEL}
\label{sec:3}

    To demonstrate this procedure, we study the quantum phase transition in a $(3+1)$-D quantum Ising model on the $\mathbb{S}^3$  manifold in this paper, which can be written as 
	\begin{eqnarray} \label{eq:Ham}
		H &=& \int d \Omega_a d \Omega_b \, U(\Omega_{ab}) [ n^0(\Omega_a)n^0(\Omega_b) - n^z(\Omega_a)n^z(\Omega_b) ] \nonumber \\
		 &&- h^x \int d \Omega \, n^x(\Omega)  - i h^z \int d \Omega \, n^z(\Omega).
	\end{eqnarray}
	Here  $\Omega = (\theta,\varphi,\chi)$  denotes the polar coordinates on $\mathbb{S}^3$ space. $n^{\alpha}(\Omega)$ is the local density operator 
	$	n^{\alpha}(\Omega) = (\hat{\psi}^{\dagger}_{\uparrow},\hat{\psi}^{\dagger}_{\downarrow}) \sigma^{\alpha} (\hat{\psi}_{\uparrow},\hat{\psi}_{\downarrow})^{T}$,
    where $\sigma^0=I_{2 \times 2}$ and $\sigma^{x,y,z}$ are Pauli matrices, and $\hat \psi(\theta,\varphi,\chi)$ is the electron operator. $U(\Omega_{ab})$ describes the repulsive interaction strength, and
    $h^x$ and $h^z$ respectively denote the transverse field and longitudinal field. 

    The first term of the Hamiltonian in Eq.(\ref{eq:Ham}) corresponds to short-ranged density-density interactions in real space, which favors spontaneous breaking of the global ($\mathbb{Z}_2$) symmetry, giving rise to two-fold degenerate ground states. The transverse field term tends to select a quantum paramagnet as the ground state, which preserves the ($\mathbb{Z}_2$) symmetry. Thus, the competition between the interaction term and transverse field term leads to a phase transition from the Ising ferromagnet to the paramagnet, and an Ising-type transition in between exists. Moreover, the longitudinal term breaks the hermitian property. It may drive a $\mathcal{PT}$ symmetry breaking transition belonging to the Yang-Lee university class \cite{PhysRev.87.404,PhysRev.87.410,PhysRevLett.40.1610}.

	Next, we rewrite the Hamiltonian in second-quantized form, with the help of the LLL projection. We express the field operator as 
    \begin{align}
    \hat{\psi}(\theta,\varphi,\chi) = \sum_{\bm{m}}{\Psi_{\bm{m}} \hat{c}_{\bm{m}}} ,
    \end{align}
    with $\Psi_{\bm{m}}$ the single-particle wavefunction on $\mathbb{S}^3$ space (see Eq.(\ref{eq:wavefun3})) and $\hat{c}_{\bm{m}}$ the annihilation operator for an $SO(4)$ Landau orbital labeled by $\bm{m}= (m_j,m_k)$. By keeping the $SO(4)$ LLL orbits, the resulting projected Hamiltonian reads
	\begin{equation}
		\begin{split}
			       H &= H_{00} + H_{zz} + H_t, \\
			H_{00} &= \sum_{\bm{m_{i=1,2,3,4}}} V_{\bm{m_1,m_2,m_3,m_4}} (\bm{c}_{\bm{m_1}}^{\dagger} \bm{c}_{\bm{m_4}})  \bm{c}_{\bm{m_2}}^{\dagger} \bm{c}_{\bm{m_3}}, \\
			H_{zz} &= \sum_{\bm{m_{i=1,2,3,4}}} V_{\bm{m_1,m_2,m_3,m_4}} (\bm{c}_{\bm{m_1}}^{\dagger} \sigma^z \bm{c}_{\bm{m_4}})  \bm{c}_{\bm{m_2}}^{\dagger} \sigma^z \bm{c}_{\bm{m_3}}, \\
			H_t &= -h^x \sum_{\bm{m}}{ \bm{c}_{\bm{m}}^{\dagger} \sigma^x \bm{c}_{\bm{m}} } -i h^z \sum_{\bm{m}}{ \bm{c}_{\bm{m}}^{\dagger} \sigma^z \bm{c}_{\bm{m}} }.
		\end{split}
		\label{eq:IsingHam}
	\end{equation}
    The matrix elements $ V_{\bm{m_1, m_2, m_3, m_4}} $, describing the electron scattering processes in the orbital space, can be expanded using Haldane pseudopotential  \cite{PhysRevLett.51.605}. 
    In Haldane's work, two $SU(2)$ spin-$s$ are projected onto the total spin-$(2s-l)$ multiplet, which can be generalized to our higher dimensional $SO(4)$ case. By scattering two fermions in representation $(I/4,I/4) \otimes (I/4,I/4) $ to the channel $(I/2-l_1,I/2-l_2)$ with $0 \le l_1,l_2 \le I/2$,  $ V_{\bm{m_1, m_2, m_3, m_4}} $ can be constructed as:
    \begin{equation}
    \begin{aligned}
		V_{\bm{m_1, m_2, m_3, m_4}} &= \sum_{l_1,l_2}V_{l_1,l_2} C_{\bm{m_1, m_2}}^{\bm{M}}C_{\bm{m_4, m_3}}^{\bm{M}} \\
		&= \sum_{l_1,l_2}V_{l_1,l_2} 
        C_{\frac I4,m_{j_1};\frac I4,{m_{j_2}}}^{J,m_{j_1}+m_{j_2}}
		C_{\frac I4,m_{k_1};\frac I4,{m_{k_2}}}^{K,m_{k_1}+m_{k_2}} \\
        & \quad \times C_{\frac I4,m_{j_4};\frac I4,{m_{j_3}}}^{J,m_{j_3}+m_{j_4}}
		C_{\frac I4,m_{k_4};\frac I4,{m_{k_3}}}^{K,m_{k_3}+m_{k_4}}
    \end{aligned}
    \label{eq:cgc}
    \end{equation}
    where $C_{j_1,m_{j_1};j_2,m_{j_2}}^{J,m_{j_1}+m_{j_2}}, C_{k_1,m_{k_1};k_2,m_{k_2}}^{K,m_{k_1}+m_{k_2}}$ are the Clebsch-Gordan coefficients, $\bm{m}$ denotes the single-particle orbit in the $ (I/4,I/4) $ representation and $\bm{M}$ denotes the $(I/2-l_1,I/2-l_2)$ representation. Here, $V_{l_1,l_2}$ is the pseudopotential defined on $\mathbb{S}^3$, characterized by two angular momentum indices, in contrast to the single-index case on $\mathbb{S}^2$. Eq.(\ref{eq:cgc}) shows the decomposition of the interaction matrix elements which guarantees the conservation of the $SU(2)_L$ and $SU(2)_R$ quantum numbers after scattering.
    In this paper, we set the pseudopotential parameters $V_{0,1}=V_{1,0}$ and vary $V_{0,0}, V_{1,0}$.  

    In this work, half-filled states are considered, in which the LLL is filled by $N=(I/2+1)^2$ electrons in total to probe the phase transitions (and the system size corresponds to the radius of the three-sphere, which scales as $r \propto N^{1/3}$).
    Without the longitudinal term,
    the Hamiltonian of this model has Ising $ \mathbb{Z}_2 $ symmetry: $ \bm{c_m} \rightarrow \sigma^x \bm{c_m} $, $SO(4)$ symmetry since all orbitals $\bm{m}$ in the LLL are in the $SO(4)$ representation $(j,k)_4 = (\frac I4,\frac I4)$, and particle-hole symmetry: $\bm{c_m} \rightarrow i \sigma^y \bm{c_m}^*, i \rightarrow -i $. While with the additional longitudinal term $-ih^zH^z$, the Hamiltonian has $\mathcal{PT}$ symmetry, which acts as $H^z \rightarrow -H^z, i\rightarrow-i$ \cite{PhysRevD.98.125003}.

	\begin{figure*}[t]
    \includegraphics[width=0.9\textwidth]{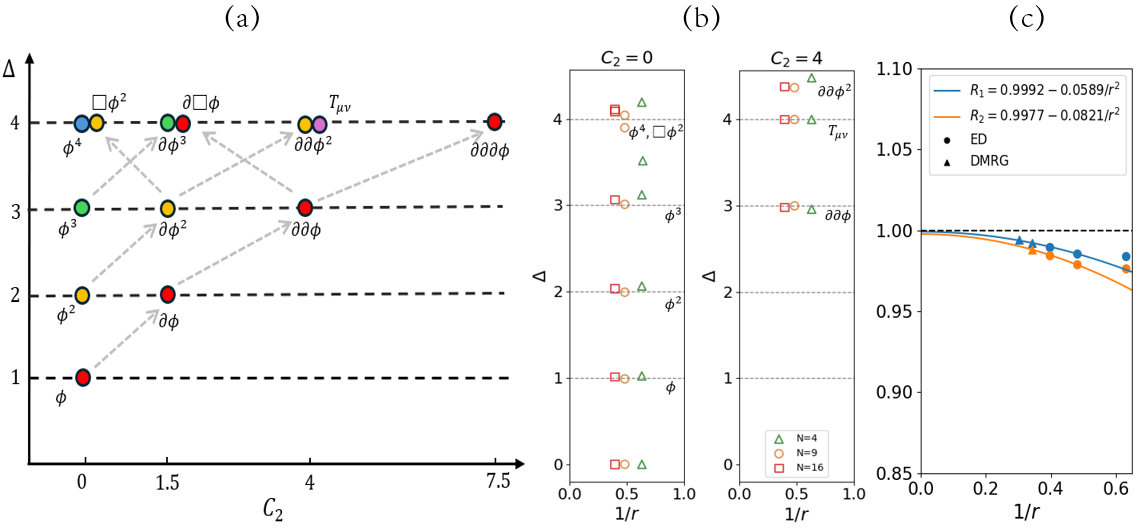} 
		\caption{ \textbf{Operator spectra for Ising transition.}
        (a) Schematic plot of the tower structure of conformal multiplets of $d=4$ free boson CFT. The grey arrow marks the operation of the conformal generator and the fields belong to the same multiplet are labeled by the same color. 
        (b) The low-lying operator spectra living in Casimir number $C_2=0$ and $4$. System sizes with $N=4,9,16$ electrons (corresponding to $SU(2)$ monopole charge $I=2,4,6$) are shown.
        (c) The ratios $R_{1,2} $ versus system sizes $1/r$ (equivalent to $SU(2)$ monopole charge $I=4,6,8,10$). Finite-size scaling gives $R_1 \to 0.9976$ and $R_2\to 0.9962$, both of which are close to the expectation $\Delta_\phi=1$. 
        $I=4,6$ are obtained by the exact diagonalization and $I=8,10$ are from the density-matrix renormalization group.
        The parameters are set to be $V_{0,0}=1.2, V_{0,1}=V_{1,0}=0.02$ and $h=h_c\approx1.3$.
        }
		\label{fig:tower}
	\end{figure*}

    \section{Numerical results}
    \label{sec:4}
	\subsection{Location of Ising Transition Point}
	\label{sec:4.1}
    We determine the Ising phase transition point $h^x_c$ of Hamiltonian Eq.(\ref{eq:IsingHam}) by using two complementary approaches.
    For simplicity, we denote $h^x$ by $h$ in the sections for the Ising transition and set $h^z=0$ in this case.
    First, we perform an optimization of the operator spectrum over the parameter space. 
    We define a cost function for quantifying the deviation of the operator spectrum from the expected values \cite{Fan:2025bhc,EliasMiro:2025msj}:
	\begin{equation}
		f(\{V\},h) =  \sum_n{(\alpha \delta E_n - \Delta_n)^2/\Delta_n},
		\label{eq:costfun}
	\end{equation} 
    where $\delta E_n$ is the numerically obtained energy gap to the $n$-th excited level, and $\Delta_n$ is the scaling dimension of the corresponding operator in the underlying Ising CFT. 
    These two quantities can be related via the state–operator correspondence.
    The Ising transition can be described in the continuum by the $\phi^4$ effective field theory. In particular, in $d=4$ its infrared fixed point is expected to be the Gaussian one, i.e., a free-boson CFT \cite{Domb_book}. The low-lying primaries include the fundamental magnetic operator $\phi$ with scaling dimension 
	$\Delta_\phi=1$, and the thermal operator $\phi^2$ with $\Delta_{\phi^2}=2$.
    In this work, the energies of the first three excitation states $\phi,\phi^2,\phi^3$ are chosen to minimize the cost function $f(\{V\},h)$ for simplicity.  $\alpha$ is the non-universal speed of light, which can often be fixed by choosing the energy-momentum tensor as $\Delta_T=4$. 
    We plot $f$ across the parameter space in Fig.\ref{fig:CPT} , and we observe a minimum of the cost function in the vicinity of $V_{0,0}=1.2, V_{0,1}=V_{1,0}=0.02$ and $h=1.3$, which is taken as the critical point in the following discussion.

    
    Second, to examine the identified  critical point, we consider the conformal perturbation theory \cite{10.21468/SciPostPhys.15.6.243,10.21468/SciPostPhys.18.3.086,10.21468/SciPostPhys.19.3.076}. We take the finite-size spectra to be realized by perturbing the CFT Hamiltonian by the relevant fields:
	\begin{equation}
		H = H_{CFT} + g_{\phi^2}\int_{S^3}{\phi^2 d^3x}+...
	\end{equation} 
	Thus, the first-order perturbation of $E_n$  can be evaluated as $E_n'(g_{\phi^2}) = \Delta_n + g_{\phi^2} \left \langle n|V_{\phi^2}|n \right \rangle $ and $V_{\phi^2}=\int_{S^3}{\phi^2d^3x}$. The cost function can be minimized over $g_{\phi^2}$ as 
	\begin{equation}
		F_{\min}(h) = \min_{g_{\phi^2}} \sum_n{(\delta E_n - E_n'(g_{\phi^2}))^2/E_n'(g_{\phi^2})}.
		\label{eq:costfun2}
	\end{equation} 
	As we tune the parameter $h$ close to the critical point $h_c$, one can expect that the optimal $g_{\phi^2}$ which minimizes the cost function should be $g_{\phi^2}=0$, since $\phi^2$ is the relevant perturbation. Therefore, we locate the critical point $h=h_c$ where $F_{\min}$ attains its minimum and the relevant perturbation $g_{\phi^2}$ crosses zero, signaling that the finite-size spectrum most closely matches the CFT expectation.
	
    As a numerical verification, Fig.\ref{fig:CPT} shows $f,F_{\min}$ as a function of transverse field $h$ for $N=9$. The term $\left \langle n|V_{\phi^2}|n \right \rangle$ used in the cost function Eq.(\ref{eq:costfun2}) is given by the perturbation theory $\left \langle n|V_{\phi^2}|n \right \rangle = C_{n \phi^2 n} r^{4-\Delta_{\phi^2}}$, where $C_{n \phi^2 n}$ is the OPE coefficient and can be constructed straightforwardly using Wick's theorem in the free $\phi^4$ theory, e.g., $C_{\phi \phi^2 \phi} = 2, C_{\phi^2 \phi^2 \phi^2}=8, C_{\phi^3 \phi^2 \phi^3}=36$. We see the function $F_{\min}$ reaches a minimum at $g_{\phi^2}=0$,  corresponding to the critical field around $h_c\approx 1.3$. Therefore, we conclude that two complementary approaches give almost the same critical field $h_c$.

	\subsection{Operator Spectra of the Ising Transition}
	\label{sec:4.2}	
	In this section, we present the numerical results for the operator spectrum, as shown in Fig.\ref{fig:tower}. All the multiplets are labeled by the eigenvalue of the $SO(4)$ Casimir operator $C_2$ and the parity of the Ising $Z_2$ symmetry. 
    In Fig.\ref{fig:tower}, we show the low-lying spectrum living in $C_2=0$ and $4$ sectors. (We cannot access the half-integer $C_2$
    since $N=(I/2+1)^2$ electrons are considered at half filling of the LLL corresponding to the $(\frac I4, \frac I4 )_4$ representation.)
    We find the numerically obtained scaling dimensions 
    are nearly stable across system sizes $N=4,9,16$ (corresponding to monopole charge numbers $I=2,4,6$). They are largely consistent with the expected operator spectra as shown in Fig.\ref{fig:tower}(a).
    The lowest field corresponds to the magnetic operator $\phi$, which is related to the local order parameter of the spontaneously broken $Z_2$ symmetry. We find its scaling dimension around $\Delta_\phi \approx 1.011$ for $N=16$, quite close to the expected value $\Delta_\phi = 1$. The second lowest field is $\phi^2$, the thermal operator driving the transition, and is estimated to have $\Delta_{\phi^2}\approx 2.033$, in agreement with the expectation $\Delta_{\phi^2}=2$. The lowest field in $C_2=4$ sector is the second descendant field of the magnetic operator, $\partial_\mu\partial_\nu \phi$.  
    For system size $N=16$, the numerical scaling dimensions of the high-lying operators [$\phi^3,\partial \partial \phi, \phi^4, \Box \phi^2$] are given by [3.065, 2.979, 4.087, 4.108], exhibiting relative errors of less than $3\%$ with respect to the CFT predictions [$3,3,4,4$].

	The integer-spacing pattern can be further verified by inspecting the ratios of different  conformal primaries. Here we define two different ratios \cite{EliasMiro:2025msj}:
    \begin{align}
      R_1&=R_{\phi,\phi^2,\phi}=(E_1-E_0) \frac{\Delta_{\phi^2}-\Delta_{\phi}}{E_2-E_1},\\
      R_2&=R_{\phi,\phi^3,\phi}=(E_1-E_0) \frac{\Delta_{\phi^3}-\Delta_{\phi}}{E_3-E_1}.
    \end{align} 
     When the system size $I \to \infty$, these ratios are expected to converge to $R_{1,2} \to 1$, since the primaries have integer scaling dimensions $\Delta_{\phi^n}=n$. In Fig.\ref{fig:tower}(c), we present $R_1,R_2$ for various system sizes ranging from $N=9-36$ ($I=4-10$), which show the convergence trend towards the expected value. 
     Then we apply a finite-size fitting process using a second-order polynomial function, which is motivated by the perturbation theory where the leading correction arises from the leading irrelevant $\phi^6$ operator.
     This gap scaling to an integer strongly suggests that the integer-spacing pattern emerges at the critical point.  

    In the numerical results, we observe that finite-size effects are evident in the high-energy part. For example,  in Fig.\ref{fig:tower}(b), we observe  a small splitting between the fields $\square \phi^2$ and $\phi^4$  that are theoretically expected to be degenerate. We attribute it to finite-size effects, as this splitting shrinks when the system size increases up to $N=16$. To further suppress finite-size effects, it would be beneficial to extend the calculation of energy spectra to larger system sizes by employing the density-matrix renormalization group (DMRG) method. We have collected some low-lying energies using DMRG calculations, which relate to $I,\phi,\phi^2,\phi^3$ at $I=8$, and $I,\phi,\phi^2$ at $I=10$. Using these data, we calculate the ratios $R_1$ and $R_2$, and we find them consistent with the expected scaling limit.

    \subsection{Yang-Lee Transition}
    \label{sec:4.3}
\begin{figure}[t]
    \includegraphics[width=0.49\textwidth]{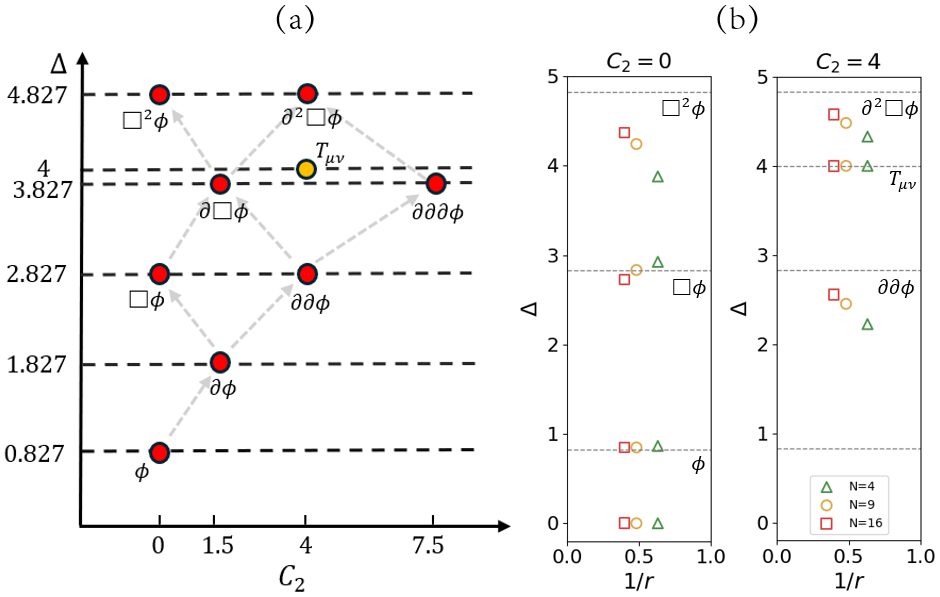} 
	\caption{ \textbf{Operator spectra for Yang-Lee transition.}
    (a) Schematic plot of the tower structure of conformal multiplets of $d=4$ Yang-Lee CFT. 
    (b) The low-lying operator spectra living in Casimir number $C_2=0$ and $4$. System sizes with $N=4,9,16$ electrons are shown.
    The parameters are set to be $V_{0,0}=0.6, V_{0,1}=V_{1,0}=0.2$ and $h^x = 3.5$.
    The critical $h^z_c$ are determined by the optimization of cost function for each system size.
    }
     
    \label{fig:LYtower}
    \end{figure}

    In this section, we turn to the Yang–Lee edge singularity\cite{PhysRev.87.410,PhysRevLett.40.1610}, driven by a finite longitudinal magnetic field $h^z$. 
    Within the same strategy shown in Sec. III A, for each system size $N$, the critical point $h^z_c$ for the Yang-Lee transition is determined by minimizing the cost function in the paramagnet phase. 
    The low-lying spectrum for the Yang-Lee transition is shown in Fig.\ref{fig:LYtower}. Different from the Ising transition, only one low-lying conformal family exists up to $\Delta < 5$. The field $\phi^2$ becomes a descendant field due to the equation of motion $\square \phi = \phi^2$ \cite{PhysRevLett.40.1610}. For $N=16$, we find the scaling dimension of the lowest primary field $\Delta_\phi \approx 0.848$.  
    In comparison, the $\epsilon-$expansion up to five-loop calculation\cite{PhysRevD.103.116024,ArguelloCruz:2025zuq} yields the estimate $\Delta_\phi \approx 0.827$, and the bootstrap gives $\Delta_\phi\approx 0.823$ \cite{10.1093/ptep/pty054}. Importantly, the descendant fields organize into an approximate integer-spacing tower structure, signaling the emergent conformal symmetry at the Yang-Lee transition.

    \section{Summary and Discussion}
    \label{sec:4}
	
	In this paper, we have explored four-dimensional (equivalent to quantum three-dimensional) phase transitions by the generalized fuzzy sphere scheme. Although the original proposal is defined on the fuzzy two-sphere $\mathbb{S}^2$, we have extended it to the fuzzy three-sphere $\mathbb{S}^3$ by implementing a projection onto the lowest SO(4) Landau level. Applying this method to the Ising transition and the Yang-Lee transition, we correctly demonstrate the conformal tower structure of low-lying critical fields, providing a clear confirmation of the underlying conformal symmetry. 

Traditionally, studying phase transitions in higher dimensions has been a computationally formidable challenge. Conventional methods, such as Monte Carlo simulations, struggle with the exponential growth of the Hilbert space, rendering precise analysis of critical behavior prohibitively expensive and often intractable for quantum systems. In stark contrast, the fuzzy sphere methodology used here establishes a novel theoretical framework that dramatically lowers the computational barrier. The essential physics governing critical points, including conformal symmetry and scaling dimensions, can be extracted through finite-size data. This low-cost and tractable approach makes the systematic exploration of rich higher-dimensional phase diagrams feasible, beyond the reach of traditional simulations.
    
    
An interesting direction for future work is the extension of this framework to higher dimensions. From a more general perspective, fuzzy spheres in higher dimensions can be systematically constructed through the lowest Landau level projection associated with non-Abelian monopole backgrounds. In particular, the underlying geometric structure is closely related to generalized Hopf maps derived from division algebra, as well as their extensions based on Clifford algebra, which provide a natural framework for constructing fuzzy $S^{2k}$ in even dimensions\cite{Hasebe:2010vp}. 
For example, one may consider generalizations based on fuzzy four-spheres\cite{doi:10.1126/science.294.5543.823, qygs-j2ys}, where the corresponding Landau level construction provides a natural finite-dimensional Hilbert space. Nevertheless, it remains an open problem to identify interacting Hamiltonians whose low-energy physics flows to nontrivial fixed points in $d>4$. We envision more studies will explore along this direction in future.


\begin{acknowledgments}
W.Z. thanks Yin-chen He for collaborating on the related projects. 
This work was supported by NSFC under No. 12474144. 
\end{acknowledgments}

\bibliographystyle{unsrt}  
\bibliography{reference}

\clearpage

\end{document}